\documentclass{aastex}
\usepackage{emulateapj5}
\usepackage{onecolfloat5}
\usepackage{apjfonts}
\usepackage{epsfig}
\usepackage{natbib}

\def\myputfigure#1#2#3#4#5%
{\vskip#5pt\makebox[0pt]{\hskip#2in
\includegraphics[width=#3\textwidth]{#1}}\vskip#4pt\hfill}

\def\gsim{\;\rlap{\lower 2.5pt
\hbox{$\sim$}}\raise 1.5pt\hbox{$>$}\;}
\def\lsim{\;\rlap{\lower 2.5pt
   \hbox{$\sim$}}\raise 1.5pt\hbox{$<$}\;}
\newcommand{\be}{\begin{equation}}
\newcommand{\beq}{\begin{equation}}
\newcommand{\ba}{\begin{eqnarray}}
\newcommand{\ee}{\end{equation}}
\newcommand{\eeq}{\end{equation}}
\newcommand{\ea}{\end{eqnarray}}

\newcommand{\nh}{$N_{\rm HI}\hspace{1mm}$}

\newcommand{\sigtot}{$\sigma_{\rm tot}(\rm{E})\hspace{1mm}$}
\newcommand{\hs}{\hspace{1mm}}
\usepackage{natbib}

\begin{document}
\twocolumn[%%% Begin front material

\submitted{Submitted to ApJ}

\title{On Detecting the X--ray Silhouette of a Damped Ly$\alpha$ System}

\author{Mark Dijkstra, Zolt\'an Haiman \& Caleb Scharf}

\affil{Department of Astronomy, Columbia University, 550 West 120th Street, 
New York, NY 10027}

\vspace{0.23cm}
\vspace{-0.5\baselineskip}

\begin{abstract}
We explore the possibility of resolving an image of a damped Ly$\alpha$ 
(DLA) system in absorption against an extended, diffuse
background X--ray source.  Typical columns of neutral hydrogen in DLA systems
are high enough to block out up to $\sim 30\%$ of the soft X--ray flux
at an observed photon energy of 0.5 keV, and we find that $\sim 1\%$
of the area of extended X--ray sources at $z \gtrsim 1$ have their 0.5
keV flux reduced by at least $20\%$ because of intervening DLA systems.  We
discuss the observability of such absorption and find that $\gtrsim 300$
photons per angular resolution element are required in the 0.3--8 keV
band for its detection, and in order to distinguish it from intrinsic
surface brightness fluctuations.  For the surface brightness of the
currently known high--redshift extended X--ray sources, this requires
an integration time of a few Msec on {\it Chandra} if the maps are
smoothed spatially to $\approx 2''$ resolution.  The exact
required integration time depends on the DLA system's column density,
metallicity and most strongly, its redshift. Current X--ray telescopes
are likely to detect DLA systems with $N_{\rm HI}<10^{22}~{\rm cm^{-2}}$ only
out to $z \approx 2.3$.  The availability of DLA systems with a suitably high
column--density for a silhouette detection is currently poorly
known. We suggest that at low redshifts, archival data of bright
X--ray point sources may be useful in constraining the high--$N_{\rm
HI}$ end of the column density distribution.  We briefly discuss an
alternative strategy of searching for extended X--ray sources behind
known DLA systems.  Although with current X--ray telescopes the detections
are challenging, they will be within the reach of a routine
observation with a next generation X--ray telescope such as {\it
the X-Ray Evolving Universe Spectrometer (XEUS)}
or {\it Generation--X}, and will deliver novel constraints on the
nature of proto--galaxies.
\end{abstract}

]

%%% End front material

%\maketitle

\section{Introduction}
\label{sec:intro}

Damped Ly$\alpha$ (DLA) absorbers owe their name to the presence
of broad damping wings in their absorption profiles caused by column
densities of neutral hydrogen in the range $N_{\rm HI}\sim
10^{20}$ to $5\times 10^{21}$cm$^{-2}$.  These high column densities are
only found in the cold disks of late-type galaxies in the local
universe, but to identify the galaxies hosting this gas at
cosmological redshifts has proven difficult, because of the
overwhelming brightness of the background quasar in the optical. Only
a dozen probable hosts for DLA systems at $z\lesssim 1$ have been identified
and their morphologies span a wide range \citep{Chen03,Rao03}.

So far several hundred DLA systems have been found in optical and ultraviolet
quasar absorption spectra in the redshift range $z\sim 0.1-4.6$
\citep{Curran02}. It has been shown that at $z\gtrsim 3$ they are the
main contributors to $\Omega_{\rm HI}$
\citep{Rao00,StorrieLombardi00,Prochaska04}.  and make up the largest
reservoir of cold gas available for starformation.  Indeed,
$\Omega_{\rm HI}$ at $z\approx 3$ is comparable to $\Omega_*$ at
$z\approx 0$, the cosmic mass density of stars in the local universe
\citep{Wolfe95}.

It has therefore been argued that DLA systems are the progenitors of
present--day galaxies.  The exact nature of these progenitors however,
is controversial: some have argued that DLA systems are large proto-galactic
disks, based on (1) the asymmetric profiles of the low-ionization
absorption lines, which are explained if the metals are present in a
large, rapidly rotating disk \citep{Prochaska97} and (2) the large
($R > 15-30$ kpc;-e.g., \citealt{Chen03}, \citealt{Chen04}) physical
separation between the absorber and possible associated starlight in a
few cases.  Others have argued that these data are consistent with
DLA systems being a collection of much smaller ($\lesssim 5 $ kpc) compact HI
clouds falling onto a galaxy in their vicinity
(\citealt{Maller01};\citealt{Haehnelt98}) and that these models
provide a better fit to the kinematic data on the high ionization gas
\citep{Maller03}. The collection of clumps in this scenario extends over
an angular scale comparable to that of the single large disk. A
distinguishing feature between the models is therefore the smoothness
of the DLA system's gas.

Determining the nature of the progenitors of our present--day galaxies
is an important step in understanding galaxy formation in general.
Obtaining a spatially resolved image of the neutral gas in a DLA system will
be an important step in settling this issue.  Unfortunately, detecting
the neutral hydrogen in DLA systems in emission at $z \gtrsim 0.5$ using the
21cm line on an existing telescope such as the Giant Metre-wave Radio
Telescope (GMRT) is not feasible (e.g. \citealt{Bagla99}).  Among
planned future radio telescopes, only the Square Kilometer Array (SKA)
has the appropriate sensitivity and frequency range to make such
observations.

A major hindrance for constraining the properties of the DLA system's gas is
the point-like character of the background quasars.  Finding DLA systems in
absorption against extended background sources would provide valuable
additional information.  In one example, \cite{Briggs89} found that
the dimensions of a disk--like absorber at $z=2$ are $> 11$ kpc
against an extended background radio source; ruling out compact gas
configurations such as dwarf galaxies. In another case,
\cite{Kanekar03} have found HI 21cm absorption in a gravitational lens
system at $z=0.76$ toward a quasar, from which the column density of
hydrogen in the lens was estimated to be $(2.6 \pm 0.3)\times 10^{21}$
cm$^{-2}$. The lens is therefore a DLA system, but because the largest
separation between a pair of radio images was small, $\sim 0.7''$
, the gas configuration and extent could not be constrained.
\cite{Fynbo99} detected extended Ly$\alpha$ emission from a DLA system,
that is illuminated by two nearby ($\Delta z \lsim$ 0.01) quasars,
and found the emission to extend over a region of $3''\times 6''$.
 A disadvantage of this technique is that it only works when
a bright source of ionizing radiation is physically very close to the
DLA system.

In this paper, we examine the complementary possibility of detecting
DLA systems against an extended background source emitting X--rays. The
motivation for this approach is the discovery of the extended X--ray
source 4C 41.17 by \cite{Scharf03} at $z=3.8$, with diffuse emission
extending over $\sim 100$ arcsec$^2$.  The X--ray spectrum is
non-thermal and has a power-law index consistent with that of radio
synchrotron emission.  \cite{Scharf03} conclude that the X--rays are
most likely inverse Compton scattered cosmic microwave background(CMB)
 or far--infrared background photons. A similar source, 3C 294, was found by
\cite{3c294} at $z=1.8$ with diffuse emission extending over $\sim
200$ arcsec$^2$.

A silhouette of a galaxy containing neutral hydrogen against a
background X--ray cluster has already been found at low redshift
($z=0.02$, \citealt{Clarke04}) and encourages us to explore diffuse
X--ray sources as possible background screens against which to image
DLA systems. \cite{Bechtold01} found strong evidence for X-ray absorption 
in the direction of a $z=0.3$ DLA system in
the spectrum of a $z=1.1$ quasar, a bright X-ray point source
with one faint jet \citep{Siem02}.

The outline of the rest of this paper is as follows.
In \S\ref{sec:model}, we introduce the X-ray optical depth through a
DLA system and calculate the fraction of the sky covered by DLA systems as a
function of their value of this optical depth.
In \S\ref{sec:method}, we describe how absorption by a DLA system  can 
theoretically be distinguished from intrinsic surface brightness fluctuations and
calculate the required integration times on {\it Chandra} to do this 
observationally.
In \S\ref{sec:discuss}, we discuss various contaminants that could mimick
absorption by a DLA system; we also address the most important model uncertainties.
The ``inverse method'' of
searching for X--ray emission behind known DLA systems is discussed in \S\ref{sec:reverse}.
Finally, in \S\ref{sec:conclusions}, we present our conclusions and
summarize the implications of this work.

Throughout this paper, we focus on an observed photon energy of
$E=0.5$ keV. This choice reflects the fact that the absorption
signature is strongest in the range $E\sim 0.3-0.6$ keV (see
\S\ref{sec:method}), and that X--ray detectors are significantly less
efficient at lower energies. We adopt the background cosmological
parameters $\Omega_m=0.27$, $\Omega_{\Lambda}=0.73$, and $h=0.71$
\citep{Spergel03}.

\section{Silhouettes of DLA Systems}
\label{sec:model}
\subsection{The X-Ray Optical Depth $\tau_{0.5}$.}
\label{sec:tau}
Typical neutral hydrogen column densities in DLA systems are in the range
$N_{\rm HI} \sim 10^{20}$ to $5 \times 10^{21}$ cm$^{-2}$ and are large
enough to block a non-negligible fraction of X--rays emitted by
sources that lie behind them. The optical depth through a column \nh\
of hydrogen atoms for a photon of energy $E$ is $\tau(E)\sim0.43$ $
(N_{\rm HI}/10^{21})(E/0.5\hs{\rm keV})^{-3.2}$, for a gas that
consists solely of a cosmic mix of $76\%$ neutral hydrogen and $24\%$
helium by mass.  Observations indicate that the mean metallicity of
DLA systems is $\sim10\%$ of the solar value with a modest redshift evolution
\citep{Prochaska2003}.  The presence of metals increases the optical
depth. For example, the cross section per hydrogen atom in a gas with
metallicity $Z=0.1 Z_{\odot}$ is $\sim 1.2(E/0.5 \hs{\rm keV})^{0.3}$
times larger than in a metal free gas (e.g. \citealt{Wilms00}).  This
shows that DLAs can show up in absorption against bright background
X--ray sources (as detected already for a point--source by
\citealt{Bechtold01}).

We define the parameter $\tau_{0.5}$ as the optical
depth through the DLA system at the photon energy $E=0.5$ keV in the observer's frame, 
$\tau_{0.5}\equiv N_{\rm HI}\sigma_{\rm tot}(0.5(1+z) \hs{\rm keV})$,
where $z$ and $N_{\rm HI}$ are the redshift and 
column density of neutral hydrogen atoms in the DLA system,
respectively, and $\sigma_{\rm tot}(E)$ is the total absorption cross--section
per hydrogen atom. This cross--section includes 
absorption by helium and heavier elements. 
To calculate \sigtot of the DLA gas, we use the model
described by \cite{Wilms00}.  This model includes up--to--date
photoionization cross sections and heavy elements in a solar abundance
pattern, as well as H$_2$ molecules.
We note that X--ray absorption arises predominantly 
from inner shells ionizations of metals, and
ionization corrections are only important at gas temperatures much
higher than those expected to occur in DLA systems.

\subsection{The Influence of DLA Metallicity on $\sigma_{tot}$ and $\tau_{0.5}$.}
\label{sec:tausig}
We consider the specific class of DLA systems with $\tau_{0.5}=0.1$.
Given \sigtot and requiring $N_{\rm HI}\sigma_{\rm tot}(E=0.5(1+z)\hs{\rm
keV})= \tau_{0.5}$, we obtain a column density of
hydrogen $N_{\rm HI}(z)$ for these DLA systems.
In Figure~\ref{fig:ncol}, we show $N_{\rm HI}(z)$ as
a function of redshift.

\myputfigure{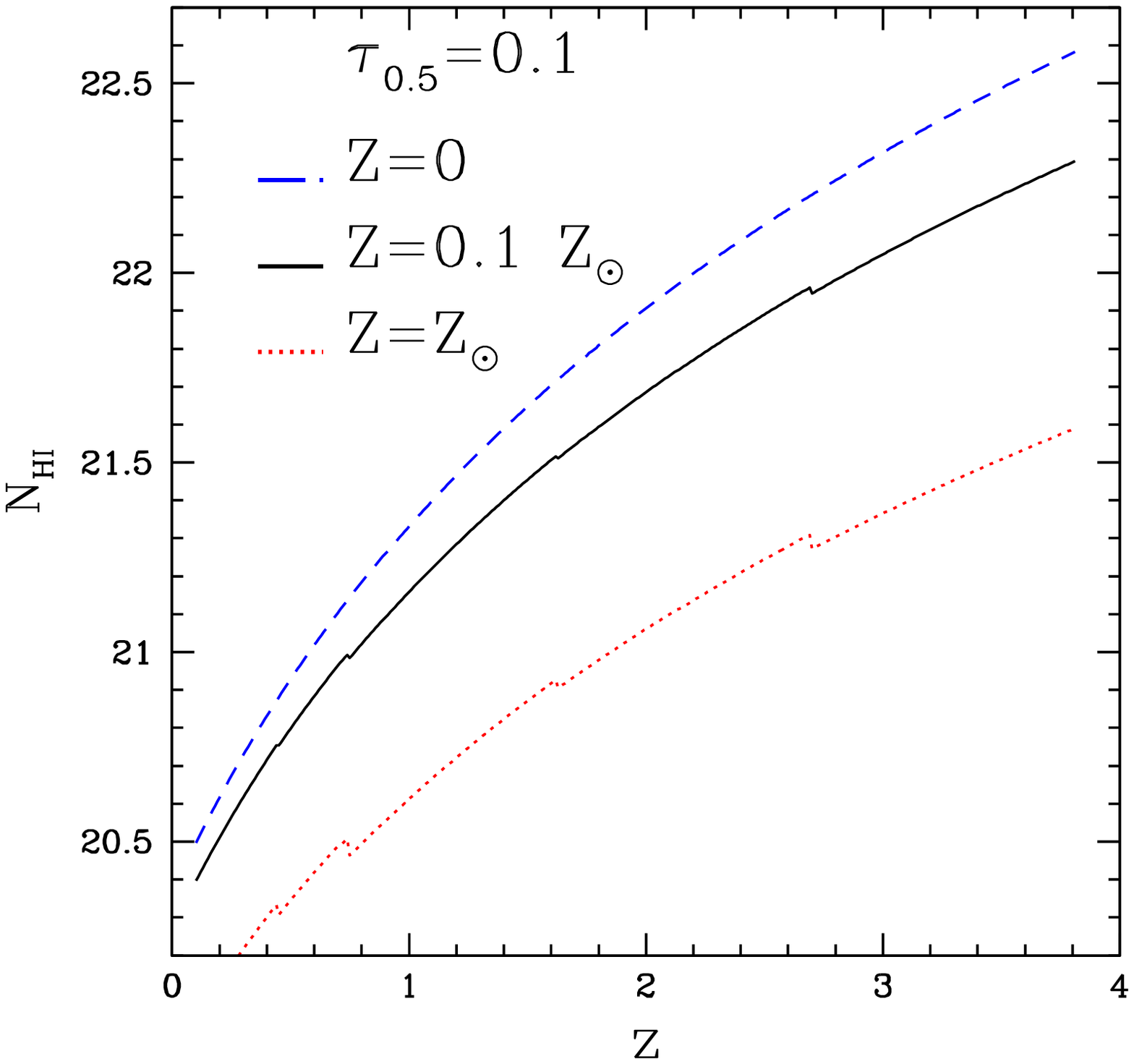}{2.8}{0.55}{-50}{-10} \figcaption{Column
density of hydrogen required to produce an absorption optical depth of
$\tau_{0.5}=0.1$ for photons with observed energy $E=0.5$ keV, as a
function of redshift for three metallicities.  To illustrate the
importance of metallicity in DLA systems, we show three cases: DLA systems with no
metals ({\it dashed--blue line}), with $Z=Z_{\odot}$ ({\it dotted--red line}) and
$Z=0.1Z_{\odot}$ ({\it solid--black line}).  The last is our fiducial
model. The highest observed column of neutral hydrogen among known DLA systems 
is log($N_{\rm HI}/{\rm cm^{-2}}$)=21.7 \citep{Curran02}.
\label{fig:ncol}}
%\vspace{1\baselineskip}
.\\
To illustrate the importance of the assumed
metallicity, we show three cases: $Z=0$ ({\it dashed--blue line}),
$Z=Z_{\odot}$ ({\it dotted--red line }), and $Z=0.1Z_{\odot}$ ({\it
solid--black line}). The last is our fiducial model, based
on the mean observed metallicity.  Note that the column of hydrogen 
required to produce a certain value of $\tau_{0.5}$ can vary by an order 
of magnitude depending on the metallicity of the gas.
The reason $N_{\rm HI}$ drops slightly but suddenly at $z\sim0.4, 0.7, 1.6, ...$ is that
$E=0.5(1+z)$ keV at these redshifts equals the K-shell ionization
energy of one of the heavy elements and $\sigma_{\rm tot}$ `jumps'
to a larger value.
For any other value of $\tau_{0.5}$, the corresponding column can be obtained by
multiplying the value taken from Figure~\ref{fig:ncol} by
$\tau_{0.5}/0.1$. We find that the following formula provides an accurate
fit to the relation between $N_{\rm HI}$, $z$, $Z$ and $\tau_{0.5}$ for a DLA system:

\begin{equation}
\tau_{0.5}=0.014 \hs A(Z)\hs \Big{(} \frac{N_{\rm HI}}{10^{21}\hs{\rm cm}^{-2}}\Big{)} 
\Big{[}\frac{0.5(1+z)}{1.5}\Big{]}^{-3.2+B(Z)},
\label{eq:fit}
\end{equation} where $A(Z)=1.0+5.39Z$, and $B(Z)=0.64\hs{\rm arctan}\big{[}\big{(} 
Z/0.19\big{)}^{0.91}\big{]}$   

This formula is good to within $5\%$ for $Z < 0.1 {\rm Z_{\odot}}$,
and to within $10\%$ for $Z < {\rm Z_{\odot}}$. The main reason for
the deteriorating accuracy of the fitting formula at larger
metallicities, is that the `jumps', in the absorption cross--section
are more pronounced in these cases. It is exactly at these jumps that
the fit is worse.

\subsection{The Covering Factor of DLAs.}
\label{sec:fcov}

In \S\ref{sec:tausig} we calculated the column of hydrogen
in a DLA system required to produce a given value of
the X-ray optical depth $\tau_{0.5}$ as a function
of its redshift. In this section we calculate what fraction
of the sky is covered by DLA systems with various minimum values
of $\tau_{0.5}$.

The total number of DLA sytems along a single line of sight (LOS) with 
$\tau_{0.5}$ or larger between redshift $0$ and $z_s$, where $z_s$ is
 the redshift of a hypothetical extended background source, is given by:

\be
\mathcal{N}=\int_0^{z_s} dz \int_{N_{\rm HI}(z)}^{\infty}dN_{\rm 
HI}f(N_{\rm HI},z),
\label{eq:fcov}
\ee where $N_{\rm HI}(z)$ is the column of hydrogen corresponding to
$\tau_{0.5}$ and $f(N_{\rm HI},z)dN_{\rm HI}dz$ is the number of DLA systems
in the range $N_{\rm HI} \pm dN_{\rm HI}/2$ and $z \pm dz/2$. We adopt
$f(N_{\rm HI},z)$ from \citet[hereafter P03]{Peroux03}, who find
that the data can be best fitted with a $\Gamma$ function used in studies
of the galaxy luminosity function.\footnote{Note that we obtain
$f(N_{\rm HI},z)$ by multiplying the functions quoted in P03 by
$dX(z)/dz$. Their function describes the number of DLA systems in the range
$X(z) \pm dX(z)/2$, where $dX(z)/dz \equiv (1+z)^2 (H_0/H(z))$ is the
absorption distance, originally introduced by \cite{Bahcall69}, where
$H_0$ and $H(z)$ are the Hubble parameters at redshift $0$ and $z$,
respectively.}  Note that $\mathcal{N}$ can equivalently be
interpreted as the two--dimensional covering factor of these DLA systems;
$f_{\rm cov}\equiv \mathcal{N}$. The exact form of $f(N_{\rm HI},z)$
is an important uncertainty in our model. The total sample of DLAs is
still too small to constrain $f(N_{\rm HI},z)$ well, especially at the low
redshifts and high--column densities that are most important for our
purposes.  We discuss this uncertainty in more detail in
\S\ref{sec:dfdz} below.

In Figure~\ref{fig:fcov}, we plot $f_{\rm cov}$ as a function of
$\tau_{0.5}$, with the X--ray source located at $z_s=3.8$ (motivated
by the source found by \citealt{Scharf03}, 4C 41.17).  The three thin
curves represent the same three cases shown in Figure \ref{fig:ncol}
in which the DLA gas has no metals ({\it dashed-blue line}),
$Z=0.1Z_{\odot}$ ({\it solid-black line}) and $Z=Z_{\odot}$ ({\it
dotted-red line}). The covering factor increases with metallicity because
the column of hydrogen required to produce a certain value of
$\tau_{0.5}$ decreases with increasing metallicity.  (The thick solid
line addresses a scatter in metallicities, as explained in
\S\ref{sec:sigfcov} below.)

Figure~\ref{fig:fcov} shows for example, that for the fiducial model ($Z=0.1Z_{\odot}$)
a fraction $f_{\rm cov}\sim 0.03$ and $0.01$ of the sky is covered by
DLA systems that absorb at least $\sim 10\%$ and $\sim 20\%$, respectively, 
of the 0.5 keV photons from the background object.  For different energies, $\tau_E\sim\tau_{0.5}$
$\times[\sigma_{\rm tot}(E)/ \sigma_{\rm tot}(0.5 \hs \rm{keV})]$,
since the absorption cross--section is close to a pure power--law.

The main conclusion from Figure~\ref{fig:fcov} is that for a source
similar to 4C 41.17, which has a size of $\sim 100$ arcsec$^2$, observed
at $\sim$arcsecond resolution, one can hope to find several pixels
covered by DLA systems with opacities up to $\tau_{0.5}\sim 0.1-0.2$.  Of
course, for a source covering a larger solid angle and/or in the case
of an unusually metal--enriched DLA system, there is a significant probability of
finding more opaque absorption. For example, in the case of an X--ray
emitting galaxy cluster with $\sim$ 10 times the area of 4C 41.17 (see
below), the figure shows that DLA systems with $\tau_{0.5}\gtrsim 0.3$ should
typically obscure several pixels.
\myputfigure{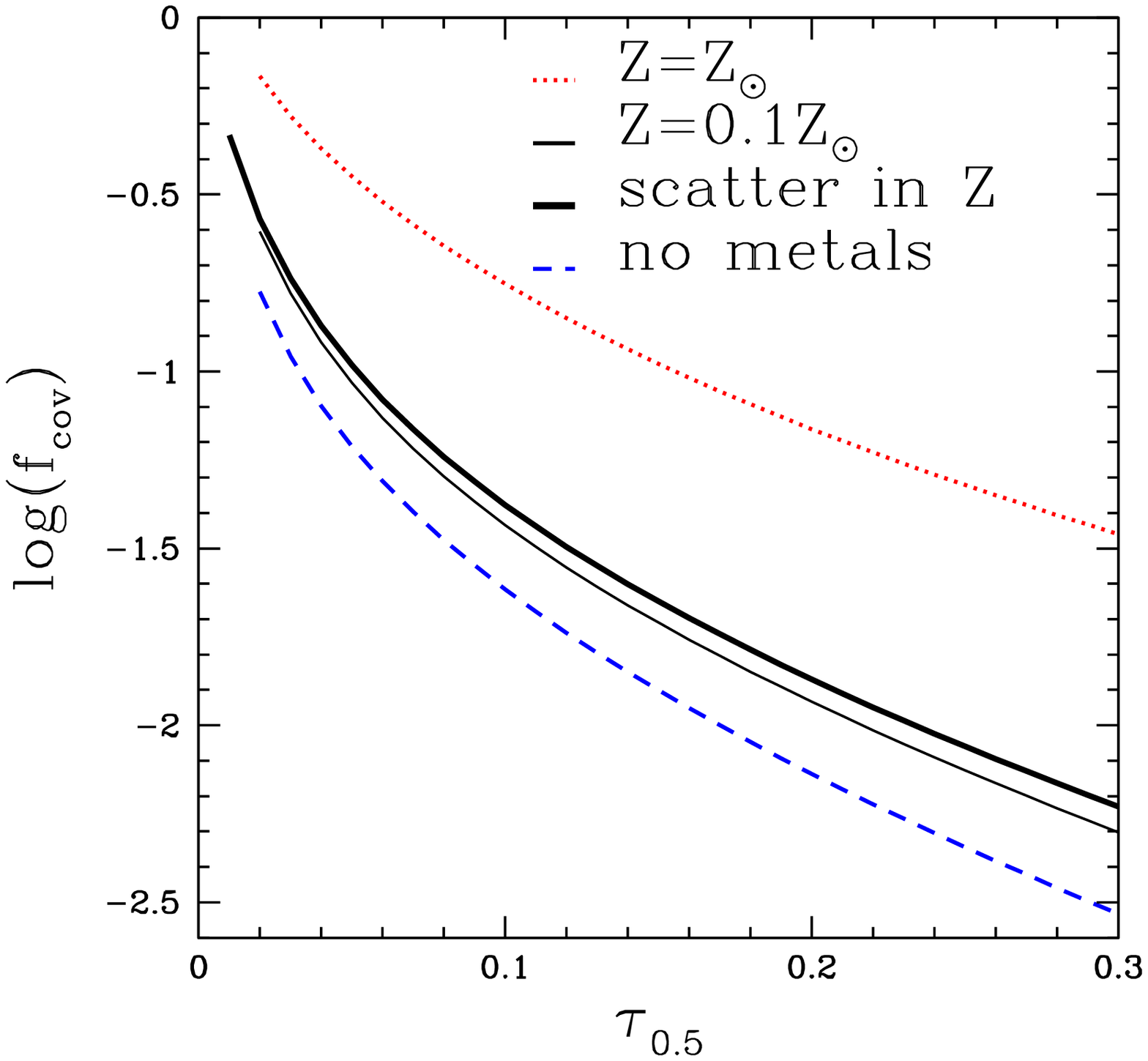}{2.8}{0.55}{-50}{-10} \figcaption{Covering
factor $f_{\rm cov}$ on the sky of DLA systems between $z=0$ and $z_s=3.8$
that have an X-ray optical depth $\tau_{0.5}$ or larger,
plotted as a function of $\tau_{0.5}$. The three curves represent the same three
metallicities as in Figure \ref{fig:ncol}. The thick, solid line
represents an alternative model, discussed in \S\ref{sec:sigfcov}, in which
the DLA systems have a log--normal distribution with mean log$(Z/{\rm Z_\odot})=-1.0$
and $\sigma_{{\rm log}Z/{\rm Z_\odot}}=0.5$.
\label{fig:fcov}}
\vspace{1\baselineskip}
\subsection{Uncertainties in $f_{\rm cov}$.}
\label{sec:sigfcov}

The calculation of $f_{\rm cov}$ has two main uncertainties. They are
discussed here.

The larger of the two uncertainties is associated with the function $f(N_{\rm HI},z)$.
The redshift of the source, $z_s$, has only a small effect on the results for
$\tau_{0.5} \gtrsim 0.1$.  The reason is that with the $f(N_{\rm
HI},z)$ adopted from P03, the absorption is dominated by low redshift
systems; e.g. $75\%$ of the contribution to $f_{\rm cov}$ comes from
$z<0.61$ and the contribution from sources with $z \gtrsim 1.4$ is
practically negligible. This inference, however, depends sensitively
on the poorly known redshift--evolution of $f(N_{\rm HI},z)$, and the
contribution to $f_{\rm cov}$ from high--$z$ systems may be much larger
(discussed further in \S\ref{sec:dfdz}).

The second uncertainty involves the metallicity of the DLA gas.
Figure \ref{fig:fcov} shows that the covering factor depends
quite strongly on the assumed metallicity. Our fiducial model
assumes $Z=0.1Z_{\odot}$, based on the mean observed value.
There is however, considerable scatter in $Z$ around this mean; 
the metallicity can easily vary by an order of magnitude between individual
objects \citep{Prochaska2003}. Since enhanced metallicity increases
$f_{\rm cov}$ more than reduced metallicity decreases it, one may expect
that incorporating the scatter enhances $f_{\rm cov}$.

We quantified in an alternative model the effect of this observed
scatter in metallicities on $f_{\rm cov}$.  A log--normal distribution
of metallicities around the mean $\rm{log}(Z_{\rm
DLA}/Z_{\odot})=-1.0$ with a standard deviation of
$\sigma_{\rm{log}(Z_{\rm DLA}/Z_{\odot})}=0.5$ was assumed (we imposed
a maximum value of $Z={\rm Z_{\odot}}$ since larger metallicities have
never been observed \citealt{Prochaska2003}).  In this model $\lesssim
10\%$ of the DLA systems have metallicities $\rm{log}(Z_{\rm
DLA}/Z_{\odot})<-1.6$ and $>-0.4$. Our results for this alternative
model are shown by the thick solid line in Figure \ref{fig:fcov}. We
find that the scatter causes a net small increase in $f_{\rm cov}$, as
expected, but that in the entire range in $\tau_{0.5}$ plotted, the
increase is $\lesssim 25\%$, which is well within the other model
uncertainty mentioned above.

\vspace{0.1cm}
\section{Observability}
\label{sec:method}

The ideal background source against which to detect a DLA system is a bright and perfectly 
smooth extended X--ray source. In this idealized case a foreground DLA system (or
that part of the DLA that intersects the extended X--ray source)
shows up in absorption against the background source in the very soft
X--ray band (0.3-0.6 keV).  In practice, extended X--ray sources are
not smooth, but have surface brightness fluctuations that are much
larger than the $\sim 10\%$ level introduced by absorption because of
intervening DLA systems. For example, the fluctuations in the number of X-ray
photons received form different locations in 4C 41.17 exceeds roughly
$100\%$. The main reason for this is small number statistics.  The mean
number of counts per pixel is $\sim$ a few, and therefore Poisson
noise will make variations between individual pixels large.  In 
3C 294 \citep{3c294}, the mean number of counts per
pixel is higher by a factor of a few and therefore Poisson noise is
less severe, although it is still large enough to introduce errors well in
excess of $10\%$. Large holes in the X-ray map of 3C 294 are clearly seen,
from which no X-rays at all are detected, and which are too large to
arise from an intervening absorber.  They probably arise from intrinsic
spatial variations in the population of relativistic particles
responsible for the scattering. We next describe two methods to distinguish the intrinsic surface
brightness fluctuations from those caused by absorption.

\myputfigure{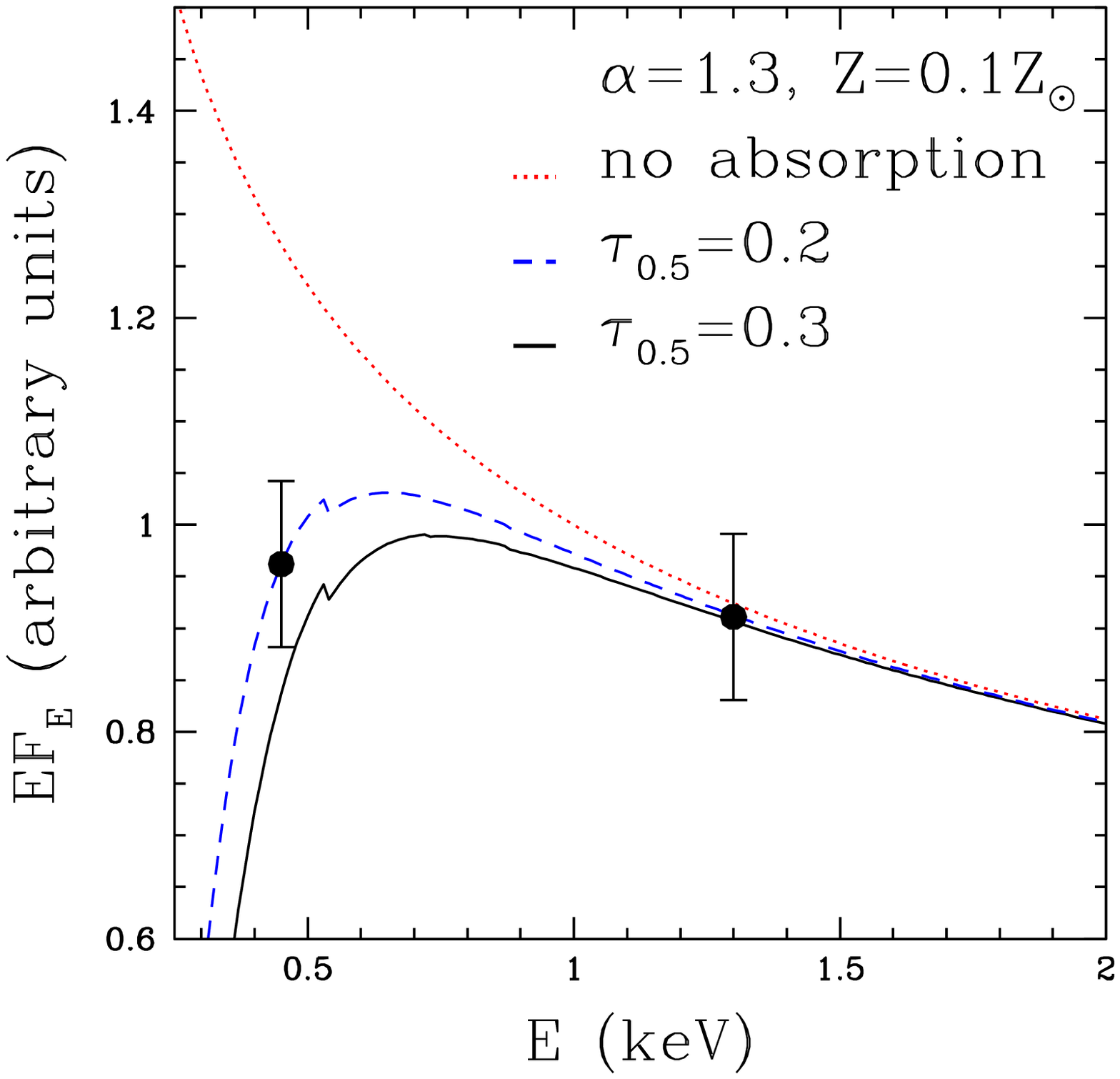}{2.8}{0.55}{-50}{-10} \figcaption{Power--law
spectrum, $F_E \propto E^{-1.3}$ of an unobscured ({\it dotted--red line}),
and an obscured X--ray source. The X-ray source is obscured by a
column of hydrogen corresponding to $\tau_{0.5}=0.2$ ({\it
dashed--blue}) and $\tau_{0.5}=0.3$ ({\it solid--black}).  In Figure
\ref{fig:ncol} the required column of hydrogen to produce
$\tau_{0.5}=0.1$ can be read off as a function of $z$. The two mock data
points show the error on the spectrum if such a source were observed
and its spectrum binned in three energy bands: 0.3-0.6, 0.6-2.0 and
2.0-8.0 keV, assuming the the total number of photons received in the
energy range 0.3-8.0 keV is $300$ (see text).
\label{fig:band}}
\vspace{1\baselineskip}

\subsection{Spectral Imprints of DLA Systems}
\label{sec:spec}

Because the absorption cross section is a strong function of energy
($\sigma_{\rm tot} \propto E^{-3}$), the spectrum of an attenuated
portion of the extended X--ray source can, in principle, reveal
whether X--rays have been absorbed by an intervening column of gas.
In order to make a crude estimate of the sensitivity of X--ray
observations to DLA attenuation, we here quantify the absorption
signature in a simple toy--model. In reality, an X--ray observatory
has a potentially complex, energy dependent, detection efficiency,
which must be taken into account in a more detailed analysis (more
discussion on this is deferred to \S\ref{sec:stat}).

In Figure~\ref{fig:band}, we show the spectrum 
of an unobscured X-ray source with a power law spectrum $F_E
\propto E^{-\alpha}$ with $\alpha=1.3$ ({\it dotted--red line
}, motivated by the inverse
Compton spectrum  observed in 4C 41.17). 
The units of the flux are arbitrary. 
Also shown is the spectrum of the same source, obscured
by an intervening column of hydrogen with $\tau_{0.5}=0.2$ 
({\it dashed--blue line}) and $\tau_{0.5}=0.3$ ({\it solid black line}).
 In these cases, absorption has removed $\sim 18\%$ and $\sim 26\%$
 of the 0.5 keV photons, respectively.
 The two mock data points with errorbars are discussed in \S\ref{sec:stat}.
Clearly, the spectra are most distinct at energies $E \lesssim 0.6$ keV. 

Unfortunately, the unobscured case is purely hypothetical
since omnipresent Galactic HI will always absorb a small
amount of X-rays. The X-ray optical depth through the Galaxy  
$\tau_{\rm Gal}$ is typically $0.2$,
although there are large parts of the
sky in which it is less than that by a factor of a few.
Spatial variations in the total column of Galactic hydrogen
across angular scales comparable to the sizes of known
extended X-ray sources are negligible. Therefore galactic
HI will obscure the entire X--ray source by the same known
amount and effectively steepen its spectrum at low energies
(see \S\ref{sec:abs}).
To find an absorption signature
of a $\tau_{0.5}=0.1$ DLA system, for example, we therefore
need to distinguish between the $\tau_{\rm Gal}=0.2$
and $\tau_{\rm Gal}+\tau_{0.5}=0.3$ cases 
(which are both shown in Fig. \ref{fig:band}).

In practice, for any source, one would proceed by first determining
the mean {\it observed} spectrum of the entire extended X--ray
source. This mean spectrum can then be used as a template, to seek
differential spectral deviations in spatial bins along the surface of
the source. 

To quantify the difference in the spectra, we introduce the ratio $\mathcal{R}$, which
is the ratio of the number of photons received in the very soft
(0.3-0.6 keV) and the soft (0.6-2.0 keV) band. In
Figure~\ref{fig:rfac}, we plot $\mathcal{R}$ in the case of
no absorption for a power law $F_E \propto E^{-\alpha}$ spectrum as a
function of $\alpha$ together with $\mathcal{R}$ for cases in
which $\tau_{0.5}=0.2,0.3$ and $0.4$. Note that the case $\tau_{0.5}=0$ 
({\it dotted--red line})
is hypothetical, but shown for completeness. The case $\tau_{0.5}=0.2$
({\it dashed--blue line})
illustrates the case in which there is Galactic absorption only, whereas
the cases $\tau_{0.5}=0.3$ ({\it solid--black line}) 
and $0.4$ ({\it long--dashed--green line}) correspond to cases in which there
is additional absorption from a DLA system with $\tau_{0.5}=0.1$ and $0.2$.

Figure~\ref{fig:rfac} shows, for example, that for $\alpha=1.3$ and
$\tau_{0.5}=0$, $\mathcal{R}=1.8$ decreases to $\sim 1.2/1.1$ and $0.9$
for $\tau_{0.5}=0.2/0.3$ and $0.4$, respectively.
There is, of course, a degeneracy with
$\alpha$; for example, $\mathcal{R}$ can also equal $0.9$ with no additional DLA
absorption when $\alpha = 0.9$. Two approaches to lift this
degeneracy are: (1) determining $\alpha$ via extrapolation from the
$E>1.0$ part of the spectrum, and/or (2) computing
$\mathcal{R}$ as a function of position, which may reveal areas where
$\mathcal{R}$ is significantly lower than in the rest of the source.
The latter approach is similar to mapping out dust using an optical
map, with the difference that obscured parts of the X--ray source look
`bluer', rather than redder.

\myputfigure{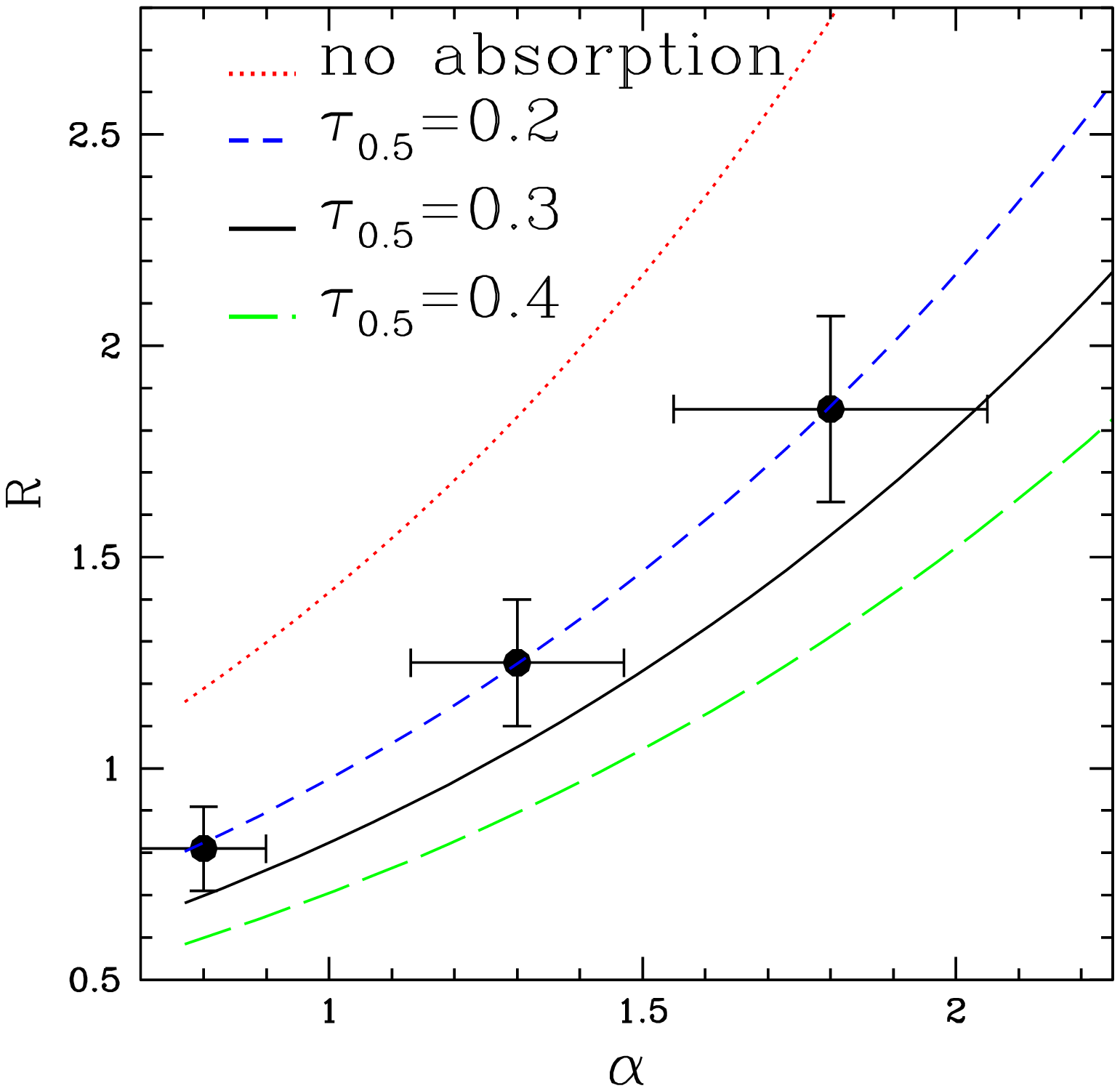}{2.8}{0.55}{-50}{-10} \figcaption{Ratio of
the number of photons in the observed 0.3-0.6 keV and 0.6-2.0 keV
bands, defined as $\mathcal{R}$, as a function of the slope of the spectrum $\alpha$. The
different curves correspond to different amounts of absorption for a
range of 0.5 keV optical depths of 0--0.4, as shown by the labels.
The three mock data points denote the error on $\mathcal{R}$ and $\alpha$
for $\alpha=0.8,1.3,$ and $1.8$, provided the total number of detected photons
in the energy range $0.3--8.0$ keV is 300.
\label{fig:rfac}}
\vspace{0\baselineskip}

\subsubsection{Photon Statistics}
\label{sec:stat}

We next address the uncertainty in $\alpha$ and $\mathcal{R}$ given that we
receive $n_{\rm ph}$ photons in the energy range 0.3-8.0 keV from a
specific location on the sky.  We bin the spectrum in the 0.6-2.0 and 
2.0-8.0 keV bands to two data points from which we extract the slope
of the spectrum, $\alpha$.  The reason we only use these two bands is
that they are not strongly affected by obscuration (see
Fig. \ref{fig:band}).

As an example, we consider the case $\alpha=1.3$:
the number of photons in the 0.3-0.6, 0.6-2.0 and 2.0-8.0 keV bands is 
149$(n_{\rm ph}/300)$, 123$(n_{\rm ph}/300)$ and 28$(n_{\rm ph}/300)$, 
respectively for $\tau_{0.5}=0.2$, which
represents the case in which there is Galactic absorption only.
The ratio of the number of photons in the 0.6-2.0 and 2.0-8.0 keV bands
is set by $\alpha$. Assuming Poisson noise on the number of photons
in each bin, we find the uncertainty in $\alpha$ and $\mathcal{R}$ to be, 
$\sigma_{\alpha} \sim 0.2$ and $\sigma_{\mathcal{R}} \sim 0.15$. 
We repeated the calculation of the uncertainty for $\alpha=0.8$ and $\alpha=1.8$.
The results for $\alpha=0.8,1.3$ and $1.8$ are shown as three 'data points' 
in Figure~\ref{fig:rfac}.
We find the following rough scaling relations of the errors: $\sigma_{\alpha} \sim 0.17(\alpha/1.3)$ and
$\sigma_{\mathcal{R}} \sim 0.15(\alpha/1.3)^{1.2}$. The $\alpha$--dependence of
the errors reflect that as $\alpha$ increases, the number of photons in the harder bands
decreases, which makes the uncertainty there larger.

Figure~\ref{fig:rfac}
shows that the $\tau_{0.5}=0.3$ and $0.4$ curves are separated by $1$ and $2\sigma$, 
respectively,from the $\tau_{0.5}=0.2$ curve, with a weak dependence on $\alpha$.
We write the level significance $S$ at which absorption can be detected as:

\begin{equation}
S=2\sigma\Big{(} \frac{\tau_{0.5}}{0.2}\Big{)} \Big{(}\frac{n_{\rm ph}}{300} \Big{)}^{0.5}
 \Big{(}\frac{1.3}{\alpha} \Big{)}^{0.5}
\label{eq:signal}
\end{equation}.

The above estimates ignore the processing of the input signal through
the X--ray instrument.  
The typical response function of X-ray telescopes like {\it Chandra}
are strong functions of energy.  {\it Chandra} is a factor of $\sim
1.5-2$ and $\sim 3-4$ more sensitive at $E=1.0$ keV than it is at $E=0.5$ and $0.3$
keV, respectively. In addition, {\it Chandra's} sensitivity decreases  roughly linearly
for $E \gtrsim 4.0$ keV.
Since the response function is known, one can correct for this, 
leaving the values of $\mathcal{R}$ we calculated unchanged. 
The errors, however, are affected by this response function.
For a fixed number of detected photons in the energy range 0.3-8.0
keV from a source with a certain $\alpha$, we find that the
total number of photons in the $0.3-0.6$ and $2.0-8.0$ keV bands is reduced
by a factor of $\sim (0.7-0.8)$ and $\sim(0.8-1.0)$, respectively
, whereas it is increased by a factor of $\sim 1.3-1.4$ in the 0.6-2.0 keV band. 
We find that this barely affects the errors on $\mathcal{R}$ and $\alpha$ and
conclude that equation~(\ref{eq:signal}) remains unchanged by a proper treatment
of the energy response function.

For comparison, \cite{Bechtold01}, detected $198$ photons the $0.4-8.0$ keV 
 energy band for a source with $\alpha=1.5 \pm 0.2$ in a 3 ks
observation of Q1331+170 with {\it Chandra}.  First we notice that the
error on $\alpha$ is the same as our estimate (note that 198 photons
in the energy band 0.4-8.0 keV, corresponds to 271 photons in the band
0.3-8.0 keV for $\alpha=1.5$).  With this number of photons, their
non--detection puts an upper limit of $\sim 10^{22}$ cm$^{-2}$ to the
column of hydrogen for the DLA system along the LOS at $z=1.77$ at
the $3\sigma$ level. Assuming $Z=0.1Z_{\odot}$, this can be translated
to an upper limit to the value of $\tau_{0.5}$ for the DLA system of 0.26. If
we interpret this upper limit of $\tau_{Gal}+\tau_{0.5}=0.46$ at the
$3-\sigma$ level as a statement that the separation between the
$\tau_{0.5}=0.46$ and $0.2$ curves is $3\sigma$, then this
is also in good agreement with equation~(\ref{eq:signal}), which
predicts $S=2.3\sigma$.  The numbers lie even closer together when we
take into account that the value of $\tau_{\rm Gal}$ in the direction of
Q1331+170 was actually $0.12$ instead of $0.2$, confirming that our
simple estimate is a good first--order approximation for the expected
signal.

\subsubsection{Required Integration Times on {\it Chandra}}
\label{sec:tint}

In \S\ref{sec:stat} we found that we need several hundred X--ray photons
to have a $2\sigma$ detection of a foreground 
DLA system along the LOS, with the exact number depending
on the slope of the X-ray spectrum of the background source,
and the value of $\tau_{0.5}$ of the foreground DLA system. To make
a map that spatially resolves a DLA system we need several hundred
photons per pixel in the image.

Ideally, we also need surface brightness maps with arcsecond resolution to
be able to differentiate between compact clumps ($d\lesssim 5$ kpc)
and large disks ($R\gtrsim 15-30$ kpc, or $D \gtrsim 30-60$ kpc,
 e.g. \citealt{Chen04}), since $1''$ corresponds to $\sim 7-8.5$ kpc 
in the redshift range 0.7-4.0.  To reduce telescope time, the maps can be binned to a
resolution of $2''\times 2''$.  If the DLA gas resides in a
large ($D \gtrsim 30-60$ kpc) disk, it will be resolved in
$\gtrsim 5-16$ resolution elements.

To obtain $\gtrsim 3\times 10^2$ photons per resolution element, we
need $\gtrsim 75$ photons arcsec$^{-2}$. We can express the required integration time
for a detection in one resolution element (pixel) at significance level $S$ as:

\begin{eqnarray}
t_{\rm int}&=3.0 \hs \Big{(} \frac{0.2}{\tau_{0.5}}\Big{)}^2
\Big{(}\frac{\alpha}{1.3} \Big{)}\Big{(} \frac{2.5 \times 10^{-5}}{{\rm counts s}^{-1}}\Big{)}
\Big{(}\frac{S}{2\sigma} \Big{)}^{2}\hs {\rm Ms}\\
&=3.0  \hs \Big{(} \frac{2.7 \times 10^{21}}{N_{\rm HI}}\Big{)}^2
\Big{(} \frac{1+z}{2}\Big{)}^{5.8}
\Big{(}\frac{\alpha}{1.3} \Big{)}\Big{(} \frac{2.5 \times 10^{-5}}{{\rm ct/s}}\Big{)}
\Big{(}\frac{S}{2\sigma} \Big{)}^{2}\hs {\rm Ms}, 
\label{eq:tint}
\end{eqnarray} where counts s$^{-1}$ is the count rate of the total number of photons
in the energy band 0.3-8.0 keV per arcsec$^{-2}$. To arrive at the second line
(eq. 5), we assumed $Z=0.1Z_{\odot}$, and used equation~(\ref{eq:fit}) to eliminate
$\tau_{0.5}$. The strong redshift dependence of the required integration time
puts constraints on which sub-classes of DLA systems can be detected in this manner.
If 3 Ms is taken as an absolute upper limit on the total integration time,
the minimum required column of hydrogen in the DLA system as a function
of redshift is $N_{\rm HI}= 2.7\times 10^{21}[(1+z)/2]^{2.9}$ cm$^{-2}$
in order for it to be detected. 
If we assume $N_{\rm HI}$ cannot exceed $10^{22}$ cm$^{-2}$, 
then the method fails to detect DLA systems at $z>2.30$.
Note that although the maximum observed log($N_{\rm HI}$ /${\rm cm}^{-2})=21.7$,
theoretically there is no evidence for an upper limit at this value of $N_{\rm HI}$.
The numbers given above are all dependent on the assumed metallicity, on $\alpha$ and
on the brightness of the background X-ray source. 

The quoted count rate of $2.5 \times 10^{-5}$ counts s$^{-1}$ is
motivated by the few recently observed high--redshift extended X-ray sources.
For example,  4C 41.17 was observed for 100 ks on {\it Chandra} by \cite{Scharf03} and has a few
photons arcsec$^{-2}$ over its $\sim 100$ arcsec$^2$ area.
Another source, 3C 294, was observed for 200 ksec with {\it Chandra}
and has $\gtrsim 4-6$ photons per arcsec$^2$ over most of its area of
$\sim 200$ arcsec$^2$ \citep{3c294}. Note that there are regions
where the count rate is higher.

\cite{Greco04} observed a distant ($z=0.8$) galaxy cluster
for 100 ksec using {\it Chandra}. In their observation an area as
large as $\sim 10^3$ arcsec$^2$ had more than $2.5$ photons 
arcsec$^{-2}$ The origin of the
X--rays in this case is mainly thermal, and the analysis given above
for a power--law inverse Compton spectrum does not apply. However, a
cluster X-ray spectrum in reality is more complex, with many heavy
element emission lines superimposed on a multi--temperature thermal
Bremsstrahlung spectrum \citep{Raymond77}.  A power law with photon
index of $\alpha\approx 0.5--1$ can therefore be used as a simple first--order
 approximation, and similar statistics will apply.
An advantage of using galaxy clusters (over the X--ray
halos of radio galaxies) is their potentially larger angular size,
which allows a search for the rarer DLA systems with more significant
opacity.

\subsection{Combining X--ray and Sunyaev--Zeldovich Effect Maps}
\label{sec:sz}
Another way to determine whether a dimming of certain part(s) in the
X--ray image is due to absorption or to an intrinsic brightness
fluctuation in the source itself is to map the CMB temperature
decrement, caused by the Sunyaev--Zeldovich (SZ) effect, along the
surface of the source. This decrement should scale with the intrinsic
X--ray brightness, regardless of whether the origin of the X--rays is
thermal or non--thermal.  In particular, thermal X--ray emission
scales with the line--of--sight integral of $n_e^2$, whereas inverse
Compton emission and the SZ surface brightness both scale with the
LOS sight integral of $n_e$ (where $n_e$ is the electron number
density).  Local density variations can therefore cause thermal
X--ray and SZ brightness variations.  However, the thermal X--ray flux is
unlikely to drop noticeably without an accompanying decrease in the SZ
decrement.  In contrast, an intervening DLA system would not have any effect
on the SZ decrement.  An SZ map with approximately arcsecond resolution would
therefore be very useful in constraining the intrinsic X--ray
brightness fluctuations. This resolution will be feasible with future
millimter telescopes, such as the {\it Atacama Large Millimeter Array}
(ALMA)\footnote{See http://www.alma.nrao.edu/}.
Whether ALMA is sensitive enough to detect SZ temperature decrements
due to sources like 4C 41.17 can be estimated as follows: As a first
order estimate, the temperature decrement at $\nu\sim100-200$ GHz can
be approximated by $\Delta T_{\rm CMB} \sim (N_e T_e/10^{28}\hs
\rm{K}\hs \rm{cm}^{-2})$ $\mu$K, where $N_e$ is the column density of
electrons in the X-ray source, $T_e=\bar{E}/k$ is the electron
temperature, and $\bar{E}$ is the mean electron energy.  The product
$N_e T_e$ is directly proportional to the observed rate of X-ray
photons, $\dot N_{\gamma}$, which is given by $\dot
N_{\gamma}=\epsilon dV/$ $(4\pi d^2_L(z) d\Omega \bar{E}_{\gamma})$
photons cm$^{-2}$ $s^{-1}$ sr$^{-1}$. Here $\epsilon=5.4 \times
10^{-36}T_eN_e(1+z)^4$ ergs s$^{-1}$ cm$^{-3}$ is the volume emissivity from
inverse Compton emission \citep[e.g.,][]{Katz96}; $d_L(z)$ is the
luminosity distance; $\bar{E}_{\gamma}$ is the mean observed photon
energy, and $dV$ and $d\Omega$ are differential volume and solid angle
elements, respectively. For 4C 41.17, $\sim 2$ photons arcsec$^{-2}$ are observed,
implying $N_eT_e\sim10^{30}$ in the above units. Therefore, we find
$\Delta T_{\rm CMB} \sim 10^2$ $\mu$K, which can be detected by ALMA
within several hours of integration \citep{Kocsis04}.

\section{Discussion}
\label{sec:discuss}

\subsection{The Absorber's Location along the LOS}
\label{sec:abs}
We found that it is possible to detect DLA systems in absorption against
background X-ray sources, but generally the redshift of the absorber
along the LOS is not constrained (except in the rare
cases in which the absorption signature is very strong;
\citealt{Bechtold01}). In this section we discuss possible
contaminants that could leave an absorption feature similar to a
DLA system along a random position along the LOS. 
The sequence in which the contaminants are discussed is based on
their distance to us.  We start with Galactic HI and end with
neutral gas physically associated with the source.

Galactic HI is present in large columns; 
the average column of hydrogen within our own Galaxy
is $\langle N_{\rm HI}{\rm sin}|b|\rangle=3\times10^{20}$ cm$^{-2}$,
where $b$ is Galactic latitude and $\langle ... \rangle$ denotes the
average over an annulus at $b$ \citealt[e.g.,][]{Lockman2003}.  Assuming
a solar abundance pattern of $Z=0.7Z_{\odot}$ (which is the 
appropriate value for the interstellar medium (ISM),
 e.g. \citealt{Wilms00}) for this column, this
translates to $\langle\tau_{\rm Gal}{\rm sin}|b|\rangle=0.19$.  Because
of spatial variations of Galactic $N_{\rm HI}$, there are large
portions of the sky where $\tau_{\rm Gal}$ is lower by a factor of a
few.  These spatial variations occur predominantly on large ($>
1^{\circ}$) angular scales, and therefore $\tau_{\rm Gal}$ can be
considered as a constant over the entire X-ray source.  The
rms spatial fractional variation in Galactic $N_{\rm HI}$ is $<3\%$
on angular scales $<1^{\circ}$, with larger fluctuations occurring on
smaller angular scales only in directions of anomalously high 
Galactic $N_{\rm HI}$ \citep{Lockman2003}. This leads to the
conclusion that the Galactic column can be treated as constant
across the extended X-ray source, and is unlikely to present a problem
beyond changing the effective slope of the X-ray source at low ($E<0.6$ keV)
energies.

The regions of anomalously high $N_{\rm HI}$ are not found only in the
Galactic plane, but clusters of HI clouds associated with the galaxy
are found at Galactic latitudes $>10-20^{\circ}$
\citep{Lockman04}. The association of this gas with the galaxy is
clear because its kinematics closely matches that of the planar gas.
Less clear is this association for the High Velocity Clouds, which have
velocities that deviate by $> 50$ km/s \citep{Wakker97} from the
allowed range based on a model of Galactic rotation. The nature of
these HVCs is not well understood, mainly because the distance to them
is observationally poorly constrained.  Possible contamination by HVCs
or extraplanar galactic clouds can be avoided simply by avoiding the
directions in which they are detected, as mentioned earlier.  However,
potentially smaller versions of HVCs or any kind of extraplanar HI
cloud may contaminate our signal.

A small compact HI cloud with an angular size of $\sim 5''$ and a 
column of a few times $10^{20}$ cm$^{-2}$ would go undetected in current
HI observations, and could mimic the absorption signature of a DLA system
at a cosmological distance.
We discuss below why these clouds, if they somehow manage to form, 
are probably short--lived and probably not a contaminant we should worry about.
For a cloud with $N_{\rm HI}=3 \times 10^{20}$ cm$^{-2}$ at a distance $D=100$ kpc
with an angular size of $\theta=5$ arcsec, we obtain an HI mass of
$\sim 10 (D/100\hs\rm{kpc})^{2}$ $(\theta/5'')^2$ M$_{\odot}$
compressed in a cloud of radius $1.2 \hs(D/100\hs
\rm{kpc})$ $(\theta/5'')$ pc. With a mean number
density of $\sim 2.5\times 10^2 (d/10\hs \rm{Mpc})^{-1}$ $(\theta/5'')^{-1}$
cm$^{-3}$ these clouds are only self--gravitating if the temperature inside them
is $\sim4(D/100\hs{\rm kpc})$ K. These cold, high density environments are typical for
molecular clouds ( although these are more massive, with $M\sim 10^5 M_{\odot}$).
Therefore one would expect these clouds to be sites of starformation as well,
in which one cloud would form only a few stars (or one star) on a short timescale.
Note that if these clouds were truly isolated, another threat to their stability
would be thermal contact with a hot surrounding medium (the Galactic halo gas
or the intergalactic medium).  Heat conduction would rapidly
raise the cloud's temperature, which would cause it to expand and lower
it's column density. 

The required columns are also typically reached in cold disks of
late--type galaxies in the nearby universe, which would be bright and
easy to identify in optical images.  Furthermore, even if the HI is associated
with faint galaxies, it will be  easier to
identify than in the case of DLAs discovered in optical quasar
spectra, because there would be no bright background quasar dominating the
optical images. 

Another possibility is that the hydrogen gas is physically associated with the source. 
Whether a significant amount of neutral hydrogen can co--exist with the hot,
X--ray emitting gas should depend primarily on the local cooling
efficiency.  In recent work, \cite{Birnboim03} and \cite{Keres04} have
argued that significant neutral gas is present only around low--mass,
high--redshift galaxies. Sources that are sufficiently X--ray bright
to allow the detection of a DLA system are likely to be associated with more
massive objects (such as powerful radio galaxies), in which the local
neutral hydrogen would be absent. In any case, absorption in a region
that is several arcseconds across would imply that the
neutral hydrogen gas associated with the source itself would extend
several tens of kiloparsecs across, which would be an interesting alternative
interpretation of any future detection.

We conclude that the two contaminants that cannot be ruled out observationally,
the compact, cold, Galactic clouds or mini HVCs and the HI associated with the 
X-ray source, both offer interesting alternative
interpretations of any future detection.

\subsection{The Uncertainties in $f(N_{\rm HI},z)$}
\label{sec:dfdz}

As mentioned above, an important uncertainty in our predictions is the
poorly known column--density distribution $f(N_{\rm HI},z)$ at low
redshifts and high columns.  The total number of DLA systems used by P03 to
constrain $f(N_{\rm HI},z)$ in the DLA range of $N_{\rm HI}$ was
$114$. These data were binned into four redshifts bins covering the redshift
ranges $z=0.0-2.0$, $2.0-2.7$, $2.7-3.5$ and $>3.5$ each containing
between 23 and 49 DLA systems.  P03 fitted a Schechter function, $f(N_{\rm
HI},z)= (f_*/N_*)$ $(N_{\rm HI}/N_*)^{-\beta}$ $\exp(-N_{\rm
HI}/N_*)$, to the data in each bin.  \cite{Prochaska04} identified six
new DLA systems at $z>3.5$ in an automated search for DLA systems in the quasar
spectra of Data Release 1 of the Sloan Digital Sky Survey (SDSS), and
found that including these six new DLA systems changed the fitting parameters
$(f_*,N_*,\beta)$ of P03 significantly at $z>3.5$ which they attribute
to the small sample size in this bin\footnote{\citet{Prochaska04} also note
that P03 have obtained systematically too low values of $N_{\rm HI}$
for log$(N_{\rm HI}/{\rm cm^{-2}})>21$.}. For the same reason the
obtained values for the fitting parameters in the lowest redshift bin
are uncertain.

Not only the fitting parameters, but also the fitting function itself is
uncertain.  Using a Schechter function to describe $f(N_{\rm HI},z)$
introduces a sharp cutoff at high $N_{\rm HI}$.  P03's $f(N_{\rm
HI},z)$ provides a good fit to the column density distribution over
most of the $z$ and $N_{\rm HI}$ range.  However, at the high-$N_{\rm
HI}$ regime that is relevant for our purposes, other functional forms
(with less sharp cut--offs in $N_{\rm HI}$) can provide equally good
fits.

We therefore find that the largest uncertainties in $f(N_{\rm HI},z)$
are in regions of $(N_{\rm HI},z)$-space that are most relevant for
our purposes. As can be seen in Figure~\ref{fig:ncol} and
equation~(\ref{eq:fit}), the column required to produce a certain
$\tau_{0.5}$ increases rapidly with redshift ($\propto (1+z)^{2.9}$,
for $Z=0.1Z_{\odot}$).  Because the high column systems are rare, the
largest contribution to $f_{\rm cov}$ in the P03 model comes from low
redshift. As mentioned in \S\ref{sec:sigfcov}, $75\%$ of the
contribution to $f_{\rm cov}$ comes from $z \lesssim 0.6$.

This dominance of low redshift DLA systems is artificial since 
the current $f(N_{\rm HI},z)$ does not capture any 
evolution in the number density of DLA systems 
in the redshift range $z=0-2$. Observationally, it is known
that the number density of DLA systems increases with redshift
between $z=0$ and $z=1$ \citep{Rao00}.
An improved determination of $f(N_{\rm HI},z)$ 
that does capture the redshift evolution in the range $z=0-2$
will weaken the dominance of low redshift systems.

In summary, our estimate of $f_{\rm cov}$ is uncertain, but is likely
to be conservative, since P03 may have underestimated the number of
high $N_{\rm HI}$ systems, especially in the highest redshift bin. In addition
, the degree to which low redshift DLA systems dominate the contribution
to $f_{\rm cov}$ is exaggerated with the current $f(N_{\rm HI},z)$.

\subsection{Using X-ray Point Sources to constrain $f(N_{\rm HI},z)$.}
\label{sec:Xpoint}

An intriguing possibility is to exploit the enhanced sensitivity to 
low--redshift systems to constrain $f(N_{\rm HI},z)$ at low redshifts.
The brightest X-ray sources \cite{Bechtold01}
have count rates of $\sim 1$ photon s$^{-1}$; $\sim 3-15$ ks integrations
on {\it Chandra} can be used to constrain 
whether and how much additional (to the Galaxy) 
absorption is present in the spectra of these sources.
X-ray sources that are fainter by factor of $\sim 10$, 
which are more common, can detect a foreground DLA system with 
$\tau_{0.5}=0.1$ at the $3\sigma$ level in 25 ksec according
to equation~(\ref{eq:tint}) (Note that for point sources we need to multiply
eq.~[\ref{eq:tint}] by 4, since in its derivation we assumed
the X--ray was extended and binned into $2'' \times 2''$ pixels.) 
The number of X-ray sources of this brightness or more
is expected to be a few tens per square degree \citep[e.g.][]{Moretti03}.
Although the redshift of the absorber is not known, with the exception
of the cases in which the absorption signal is very strong \citep{Bechtold01},
it is constrained by the redshift of the source. 

It therefore seems promising to look at archival data in which sources
brighter than $0.1$ counts s$^{-1}$ have been observed for $25$ ks, or
more generally, any archival data that has $\gtrsim 2.5 \times
10^3(S/3\sigma)^2$ counts from a given location and see if any
additional absorption is detected at the level S. The main difficulty
with this method is that the X-ray point source itself may be
obscured.  As a result, the amount of absorption attributed to any DLA system
will be uncertain.  Statistically, the distribution $f(N_{\rm HI},z)$
of DLA systems could still be inferred, but would require modeling of the
intrinsic obscuration (e.g. \cite{Gilli01}). A more robust approach
would be to select unobscured sources, and derive upper limits on any
DLA absorption, which would constrain (put upper limits) on $f(N_{\rm
HI},z)$.

\subsection{Constraints on Galaxy Formation Scenarios}
\label{sec:zens}

As mentioned in \S~\ref{sec:intro}, a spatially resolved map of the
DLA gas could distinguish between the two models in which the gas resides
either in a large disk, or in smaller cold clumps accreting onto to
the galaxy.  The large--disk models apply at $z>2$, and as we saw in
\S\ref{sec:fcov} the contribution to $f_{\rm cov}$ form $z>1.4$ is
negligible. Although we found in \S\ref{sec:dfdz} that this dominance
of the lower redshift DLA systems is exaggerated, we also found in
\S\ref{sec:tint} that long, $\gtrsim 3$ Ms integration times are
required to detect DLA systems at $z>2.3$ (eq.~\ref{eq:tint}). Unless future
X-ray observations reveal extended X-ray sources with count rates $>
2.5\times 10^{-5}$ counts s$^{-1}$, we will not be able to detect $z > 2.3$ DLA systems
in absorption with log$(N_{\rm HI}/{\rm cm}^{-2})\lsim 22$, which
would exclude all currently known DLA systems.

We conclude that current X-ray telescopes most likely will not be able to 
produce maps that would enable us to distinguish between
the distributions of gas predicted by the different galaxy formation
scenarios. This is even more so, because the integration times mentioned
above are for X-ray maps smoothed to $2''\times 2''$.
At $z=2$, $2''$ corresponds to $\sim 16$ kpc, and small clumps would not
be resolved. It is, however, possible to produce maps of DLA gas at $z<2$. 
Knowledge of the $z<1$ systems is limited, because
currently only 23 are known. For these 23 DLA systems, 14 have confirmed
galaxies as counterparts, which are consistent with being drawn from a field
population of galaxies \citep[e.g.][]{Chen04}. For a few of these galaxies with a stellar disk,
the rotation curve of the ionized gas within the stellar part of the galaxy
was compared with the velocity difference between the galaxy and the DLA gas.
Based on the results and the distance of the DLA gas to the center of the galaxy,
it was found that the gas in the $z<1$ systems was consistent with being
in a large disk, extending to radii larger by a factor of $\sqrt{2}$
than those observed in local disk galaxies (see \citealt{Chen04} and references therein).
Producing maps of the DLA gas in these $z<1$ systems may provide an important test for this result.
Increased knowledge of the $z<2$ systems will clearly add to the understanding of their higher redshift counterparts.

\section{Mapping out Known DLA Systems}
\label{sec:reverse}

So far we have focused on examining an extended X--ray source and
searching for evidence of an intervening DLA system. As mentioned above,
several hundred DLA systems are known to exist in the redshift range
$z=0.1-4.6$.  A complementary approach would be to look for extended
X--ray sources behind these DLA systems, and examine these extended
sources for evidence of extended absorption from the DLA system.

\cite{Scharf03} choose 4C 41.17 for X--ray study because it was an
example of a cluster of submillimeter sources, centered on a powerful radio
galaxy, and was therefore considered a probable proto--galaxy--cluster
with X--ray emission.  Extended X--ray sources at high redshifts may
indeed turn out to be common, especially around radio--loud galaxies.
Of the several hundred known DLA systems in front of quasars, $\sim 10\%$
w be along sight--lines to radio loud quasars. A certain fraction
of these radio loud quasars will be similar to 4C 41.17. This fraction
is unknown at present, but is unlikely to be negligible.

The first question is, what fraction of the known DLA systems have
$\tau_{0.5}\gtrsim 0.1$, and would therefore be good candidates for
such a study.  A table containing $\sim 300$ DLA systems is given in
\cite{Curran02}, for which we calculated the corresponding values of
$\tau_{0.5}$. Since the metallicity is unknown for most of them, we
assumed it to be $0.1Z_{\odot}$.  We found a large spread in the
values of $\tau_{0.5}$, with the maximum as high as $\tau_{0.5}=1.2$.
We also found that 12 and 7 DLA systems have $\tau_{0.5}>0.10$ and $>0.20$, respectively
 (while this number was reduced to 8/6 when we lowered the metallicity to
$0.01Z_{\odot}$). The DLA systems with $\tau_{0.5}>0.20$ all lie at $z<0.7$,
and there are three at $z>1.4$ with $\tau_{0.5}=0.10-0.12$.  The
highest value of $\tau_{0.5}=1.2$ is for the source detected by
\cite{Bechtold01}, already mentioned above.  The spectral signature is
so evident in this case, that, based on the amount of X-ray absorption
and the known HI column in the DLA system from its radio absorption
signature, the metallicity of the DLA gas could be derived, and was
found to be $\sim 0.23 Z_{\odot}$.

The second question is: what fraction of these relatively opaque
absorbers have extended X--ray emission surrounding the optical
quasars in which they were discovered?  An indicator for the presence
of extended X--ray emission may be radio loudness (powerful radio
galaxies may drive such X--ray emission, such as that found in the case of
4C 41.17).  The table of DLA systems composed by \cite{Curran02} lists the
radio flux densities of the background quasars.  We find that among
the 12 DLA systems with $\tau_{0.5}\gtrsim 0.1$, at least three are in front
of either a 3C or 4C source, which are radio loud. Among the other 8,
at least 6 have detections in the radio.  An intriguing example is
Q1354+258, which is similar to 4C 41.17:
it appears to have an extended structure in the radio, and is
additionally known to have a diffuse Ly$\alpha$ blob around it
\citep{Heckman91}. This source would be a good first candidate
to look for accompanying extended X--rays.  
The DLA system in front of Q1354+258 is located at
$z=1.42$, and it has a column of $3.2 \times 10^{21}~{\rm cm^{-2}}$
\citep[e.g.,][]{Rao00}. Note that the metallicity of this DLA system is $Z
\sim 0.02 Z_{\odot}$ \citep{Vladilo02} instead of $Z=0.1Z_{\odot}$ which slightly
lowers $\tau_{0.5}$ from $0.12$ to $0.09$. This implies that
if extended X-rays are detected from Q1354+258, it should be brighter
than 4C 41.17 for a detection at the $S=2-\sigma$ level in $\lsim 3$ Ms
on {\it Chandra}.

The above exploratory remarks suggest that a full systematic search
for DLA systems, whose parent quasars are similar to 4C 41.17, and deep X--ray
observations of these sources, is a promising alternative approach to
discover X--ray silhouettes of DLA systems.

\section{Conclusions}
\label{sec:conclusions}

The neutral hydrogen column densities in observed DLA systems are in the
range log$(N_{\rm HI}/{\rm cm^{-2}})=20-21.7$
(e.g. \citealt{Curran02}) and are high enough to absorb a measurable
fraction of soft X--ray photons from a background source.  We defined
the parameter $\tau_{0.5}$, which is the optical depth through DLA systems
for a photon with an observed energy of 0.5 keV.  We showed in \S\ref{sec:fcov}
that $\sim 1\%$ of the sky is covered by DLA systems
 which have $\tau_{0.5}\geq0.2$ when
the background X--ray sources lie at redshifts $z \gsim
1.4$. This result depends on the assumed column density distribution
$f(N_{\rm HI},z)$, which is at present poorly constrained, and also on
the mean metallicity of the DLA systems, which we assumed to be
$Z=0.1Z_{\odot}$ based on the observed value
(e.g. \citealt{Prochaska2003}). We discussed uncertainties
in $f_{\rm cov}$ in \S\ref{sec:sigfcov} and found that the 
values we found are most likely conservative.

An optical depth of $\tau_{0.5}=0.2$ will introduce a dimming in the
soft X--ray bands by $\sim 20\%$,  an amount that is much lower
than intrinsic brightness fluctuations. However, as we discussed in
\S\ref{sec:method}, absorption by an intervening DLA system can be
distinguished from intrinsic brightness fluctuations by looking for
the spectral signatures of absorption (\S\ref{sec:spec} ,
Fig. \ref{fig:band}) and/or combining the X--ray observations with an
SZ decrement map of the area (\S\ref{sec:sz}).
We showed in \S\ref{sec:stat} that in order to reliably detect a DLA 
absorption signature on top of the spectral imprint left by Galactic
HI, $\gtrsim 300$ photons per angular resolution element are required
in the 0.3--8 keV band to detect a DLA system with $\tau_{0.5}=0.2$ at the
$2\sigma$ level.

For diffuse, high redshift X--ray sources similar to 4C 41.17
\citep{Scharf03} and 3C 294 \citep{3c294}, this requires observation
times of a few megaseconds on {\it Chandra} when the maps are smoothed
to $\sim 2''$ resolution. The influence of the brightness 
of the background X-ray source, the slope of the X-ray spectrum and
the value of $\tau_{0.5}$ on the required integration time on {\it Chandra}
are detailed in \S\ref{sec:tint}.

We discuss the implications of the poor constraints on the absorber's
redshift using this method in \S\ref{sec:abs}
and find two possible contaminants that cannot be ruled out observationally.
These are (1) neutral hydrogen physically associated with the X-ray source and
(2) a small, compact nearby HI cloud. Both are astrophysically interesting as we
argued in \S\ref{sec:abs}.
We suggested a new method to use archival data of bright
(count rates > 0.1 photons s$^{-1}$) X-ray point sources
to constrain the poorly known low-redshift column density distribution function,
$f(N_{\rm HI},z)$, in \S\ref{sec:Xpoint}

The inverse method is discussed in \S\ref{sec:reverse} in which we
presented the possibilities of mapping out known DLA systems using an X-ray
background source. We found that out of the sample of $\sim 300$ DLA systems
tabulated in \cite{Curran02}, $12$ have $\tau_{0.5}>0.1$, with values
anywhere in the range $\tau_{0.5}=0.1-1.2$.  We found that at least 4
of these sources resemble 4C 41.17 and 3C 294 in their radio loudness
and at least one of them also has a surrounding extended Ly$\alpha
$ halo. These 4 sources are promising candidates to have extended
X-ray emission associated with them, and, if such X--ray emission is
revealed in deep X--ray follow--ups, they would be suitable to search
for DLA silhouettes.

Although challenging, X--ray silhouettes are likely to be the only
viable method to directly spatially resolve hydrogen gas in distant
DLA systems in the foreseeable future.  Among planned future radio
telescopes, only SKA will be able to detect 21cm emission from neutral
hydrogen at $z>0.5$ \citep[e.g.,][]{Hulst2004}, and extended
Ly$\alpha$ emission would only be detectable for DLAs in the
immediate vicinity of a luminous ionizing source.

The planned X--ray observatories, {\it the X--ray Evolving Universe Spectrometer}
({\it XEUS}; \citealt{Parmar03}) and {\it
Generation X} \citep{Zhang2002} are expected to have $\sim 0.1''-1''$
resolution, and a sensitivity of 1 or 2 orders of magnitude
better than {\it Chandra}, extending to energies as low as 0.1 keV. At
these low energies, absorption by a foreground DLA system will leave a much
stronger imprint.  The X--ray silhouettes of DLA systems should be feasible
to detect in routine observations with these planned instruments.

\acknowledgements We thank Joop Schaye, Jason Prochaska and Art Wolfe
for useful comments. We thank the referee, Ariyeh Maller, for
constructive comments that significantly improved the presentation of this paper.
ZH gratefully acknowledges support by the
National Science Foundation through grants AST-0307291 and AST-0307200
and by NASA through grants  HST-GO-09793.18 and NNG04GI88G.
CS acknowledges the support of NASA/ {\it Chandra} grant SAO G03-4158A.

\vspace{-2\baselineskip}

\bibliography{adssample}

\end{document}